\documentclass[twocolumn,showpacs,showkeys,preprintnumbers,amsmath,amssymb]{revtex4-1}

\usepackage{graphicx}
\usepackage{dcolumn}
\usepackage{bm}

\begin{document}

\preprint{Submitted to Phys.\ Plasmas}  

\title[Vortices by HAR waves]{Vortices at the magnetic equator generated by \\ hybrid Alfv$\acute{\bf e}$n resonant waves}

\author{Yasutaka Hiraki}
 \email{hiraki.yasutaka@nipr.ac.jp}
\affiliation{
National Institute of Polar Research, Midori-cho 10-3, Tachikawa, Tokyo, 190-8518, Japan.\\
}


\date{\today}

\begin{abstract}
We performed three-dimensional magnetohydrodynamic simulations of shear Alfv$\acute{\rm e}$n waves in a full field line system with magnetosphere-ionosphere coupling and plasma non-uniformities. 
Feedback instability of the Alfv$\acute{\rm e}$n resonant modes showed various nonlinear features under the field line cavities: i) a secondary flow shear instability occurs at the magnetic equator, ii) trapping of the ionospheric Alfv$\acute{\rm e}$n resonant modes facilitates deformation of field-aligned current structures, and iii) hybrid Alfv$\acute{\rm e}$n resonant modes grow to cause vortices and magnetic oscillations around the magnetic equator. 
Essential features in the initial brightening of auroral arc at substorm onsets could be explained by the dynamics of Alfv$\acute{\rm e}$n resonant modes, which are the nature of the field line system responding to a background rapid change. 
\end{abstract}

\maketitle

\section{Introduction}\label{sec: 1}
It has been debated that Alfv$\acute{\rm e}$n field-line resonances are directly related to various phenomena occurring at the auroral region, such as auroral activation and deformation [Samson et al., 1992; Lysak and Song, 2008; Rae et al., 2009], ionospheric density disturbances [Rankin et al., 2004], auroral particle acceleration and turbulences [G$\acute{\rm e}$not et al., 2000; Chaston et al., 2003], and high-$\beta$ plasma instability in the magnetosphere [Erickson et al., 2000]. A steep cavity of the Alfv$\acute{\rm e}$n velocity has been known to exist below a height of $\approx 6000$ km, where so-called ionospheric Alfv$\acute{\rm e}$n resonant  (IAR) waves are excited [e.g., Carlson et al., 1998]. The IAR waves are distributed in a high frequency range of 1--10 Hz [Hebden et al., 2005]. On the other hand, there are global field-line resonant modes in a low frequency range of 10--100 mHz [e.g., Rae et al. 2009] that lies in a slower cavity at the magnetospheric side. These two work on a low-$\beta$ plasma medium with magnetosphere-ionosphere (MI) coupling. 

Generation mechanisms of the inertial Alfv$\acute{\rm e}$n waves with a field-line wavelength of $<1000$ km and a perpendicular one of $<1$ km at the ionospheric cavity region have been vigorously studied by a particle-in-cell simulation [G$\acute{\rm e}$not et al., 2004] and satellite observations [Chaston et al., 2006]. By assuming the Alfv$\acute{\rm e}$n velocity as $v_{\rm A}\approx 1000$ km/s, they roughly correspond to the above IAR modes with frequency of $>1$ Hz; similarly, waves with $<100$ mHz correspond to a long wavelength range of $>10^4$ km. Accompanied by auroral activation at the substorm onset, magnetic pulsations (20--80 s) called Pi 1 and 2 [Holter et al., 1995; Park et al., 2012] and auroral kilometric electromagnetic radiations (AKR) [Morioka et al., 2010] have been known to occur. These phenomena are good implications for the existence of the above field line resonant modes. The IAR wave trapping can be followed by an enhancement of field-aligned currents, which leads to excitation of the inertial Alfv$\acute{\rm e}$n waves and electron acceleration. 

How much knowledge we get about generation of the IAR waves and inertial Alfv$\acute{\rm e}$n waves from the viewpoint of natural conditioning? Is it the unique solution that an impulse propagates from the magnetosphere and some waves are trapped in the ionospheric cavity, although the related auroral arc slowly develops? We have made a series of studies on the characteristics of the Alfv$\acute{\rm e}$n resonant modes destabilized in a full field line system with MI coupling. A three-dimensional magnetohydrodynamic (MHD) simulation demonstrated that the feedback instability of field line resonant modes induces vortex structures at the magnetic equator [Watanabe, 2010]; the system is a slab magnetic field and $v_{\rm A} = {\rm const}$. Considering a dipole magnetic field and steep gradients of $v_{\rm A}$, a linear analysis revealed that low-frequency modes [HW, 2011], IAR high-frequency modes, and hybrid Alfv$\acute{\rm e}$n resonant (HAR) modes [HW, 2012] become feedback unstable; the third mode is generated by a coupling of the former two. If one includes the other, or, if the IAR and HAR modes naturally grow in the process of nonlinear evolution of the low-frequency modes, a problem could be unified: the relationship between the enhancement of global convections and auroral particle acceleration along with turbulent structures. 

The purpose of this study is to comprehend the nonlinear evolution of Alfv$\acute{\rm e}$n waves underlying phenomena related to auroral intensification such as Pi 1 and 2, AKR, and electron acceleration. We clarify how the position and timing of wave growth change in response to changes in the $v_{\rm A}$ cavities. The main points are summarized below. i) On the time lag of wave amplification at the ionosphere and the magnetic equator. In case of $v_{\rm A}={\rm const}$, although the primary (feedback) instability is triggered by the MI coupling region, a secondary increase in flow and current rapidly starts at the magnetic equator side. We clarified this mechanism in a quantitative analysis. Considering the ionospheric cavity, we found that the IAR short-wavelength wave is trapped through the growth of a long-wavelength wave. ii) On the formation of an energy transport pass from the ionosphere to the magnetic equator. Considering both the ionospheric and magnetospheric cavities, we found that the IAR wave is trapped, and then part of the wave propagates to the magnetic equator side, and a flow shear instability occurs due to excitation of the HAR waves.

\section{Model Description}\label{sec: 2}
In order to elucidate the physics involved in auroral structures, nonlinear evolution of shear Alfv$\acute{\rm e}$n waves propagating along the dipole magnetic field $\mbox{\boldmath $B$}_0$ can be modeled by using two-field reduced MHD equations [Chmyrev et al., 1988]. The waves slightly slip ($\Omega / k_\perp \ll v_{\rm A}$) through the feedback coupling to density waves at the ionosphere, where $\Omega$: frequency, $k_\perp$: perpendicular wavenumber, and $v_{\rm A}$: Alfv$\acute{\rm e}$n velocity. As the model formulation in this paper is about the same as Hiraki [2014], we simply explain the coordinates, equations, and numerical techniques, while the different parts in detail. The system of interest is a field line with a length of $l\approx 7\times 10^4$ km and at a latitude of 70$^\circ$, where auroral arcs develop. The field line position $s$ is defined as $s=0$ at the ionosphere and $s=l$ at the magnetic equator (a radial distance of $\approx8.5$ Earth radius). We set a local flux tube: a square of ($l_\perp \times l_\perp$) with $l_\perp = 10^{-3} l \approx 70$ km at $s=0$, a rectangle of ($h_\nu l_\perp \times h_\varphi l_\perp$) at $s$, and ($\approx 3300$ km $\times$ $\approx 1700$ km) at $s=l$ using dipole metrics $h_\nu(s)$ and $h_\varphi(s)$ with $B_0(s) = 1/h_\nu h_\varphi$ [HW, 2011]. 

The electric field $\mbox{\boldmath $E$}$ is partitioned into a background convective part $\mbox{\boldmath $E$}_0$ ($\perp \mbox{\boldmath $B$}_0$) and the Alfv$\acute{\rm e}$nic perturbation $\mbox{\boldmath $E$}_1 = B_0 \mbox{\boldmath $\nabla $}_\perp \phi$. The magnetic perturbation is expressed as $\mbox{\boldmath $B$}_1 = \mbox{\boldmath $\nabla $}_\perp \psi \times \mbox{\boldmath $B$}_0$. The equations at $0<s \le l$ are written as 
\begin{eqnarray}
 \displaystyle && \partial_t \omega + \mbox{\boldmath $v$}_\perp \cdot \mbox{\boldmath $\nabla$}_\perp \omega = v_{\rm A}^2 \nabla_\parallel j_\parallel  \\
 && \partial_t \psi + \mbox{\boldmath $v$}_0 \cdot \mbox{\boldmath $\nabla$}_\perp \psi + \frac{1}{B_0} \nabla_\parallel B_0 \phi = -\eta j_\parallel. 
\end{eqnarray}
The convective drift velocity $\mbox{\boldmath $v$}_0=\mbox{\boldmath $E$}_0 \times \mbox{\boldmath $B$}_0/B_0^2$ is set so that $E_0$ satisfies the equi-potential condition, while $\mbox{\boldmath $v$}_\perp = \mbox{\boldmath $v$}_0 + \mbox{\boldmath $v$}_1(\mbox{\boldmath $E$}_1)$, vorticity $\omega = \nabla_\perp^2 \phi$, field-aligned current $j_\parallel = - \nabla_\perp^2 \psi$, and $\nabla_\parallel = \partial_s + \mbox{\boldmath $b$}_0 \cdot \mbox{\boldmath $\nabla $}_\perp \times \mbox{\boldmath $\nabla $}_\perp \psi$. 

Ionospheric plasma motion including density waves is described by the two fluid equations. Considering the current dynamo layer (height of 100--150 km), we can assume that ions and electrons respectively yield the Pedersen drift $\mbox{\boldmath $v$}_{\rm i} = \mu_{\rm P} \mbox{\boldmath $E$}-D \mbox{\boldmath $\nabla $}_\perp \ln n_{\rm i}$ and the Hall drift $\mbox{\boldmath $v$}_{\rm e} = \mu_{\rm H} \mbox{\boldmath $E$} \times \mbox{\boldmath $B$}_0/B_0 - \mbox{\boldmath $j$}_\parallel/e n_{\rm e}$, with $\mu_{\rm P,H}$: mobilities and $D$: diffusion coefficient. By integrating the continuity equations over the dynamo layer, the equations at $s=0$ become 
\begin{eqnarray}
 \displaystyle && \partial_t n_{\rm e} + \mbox{\boldmath $v$}_\perp \cdot \mbox{\boldmath $\nabla$}_\perp n_{\rm e} = j_\parallel - R n_{\rm e} \\
 && \mbox{\boldmath $\nabla$}_\perp \cdot (n_{\rm e} \mu_{\rm P} \mbox{\boldmath $E$}) - \mbox{\boldmath $v$}_\perp \cdot \mbox{\boldmath $\nabla$}_\perp n_{\rm e} = D \nabla_\perp^2 n_{\rm e} - j_\parallel, 
\end{eqnarray}
where $R n_{\rm e}$ is a linearized recombination term, and the Hall mobility is normalized to be unity. We assume that $j_\parallel$ is carried by thermal electrons. 

\begin{figure}[t!]
\includegraphics[width=1.0\columnwidth, bb=0 0 360 252, clip]{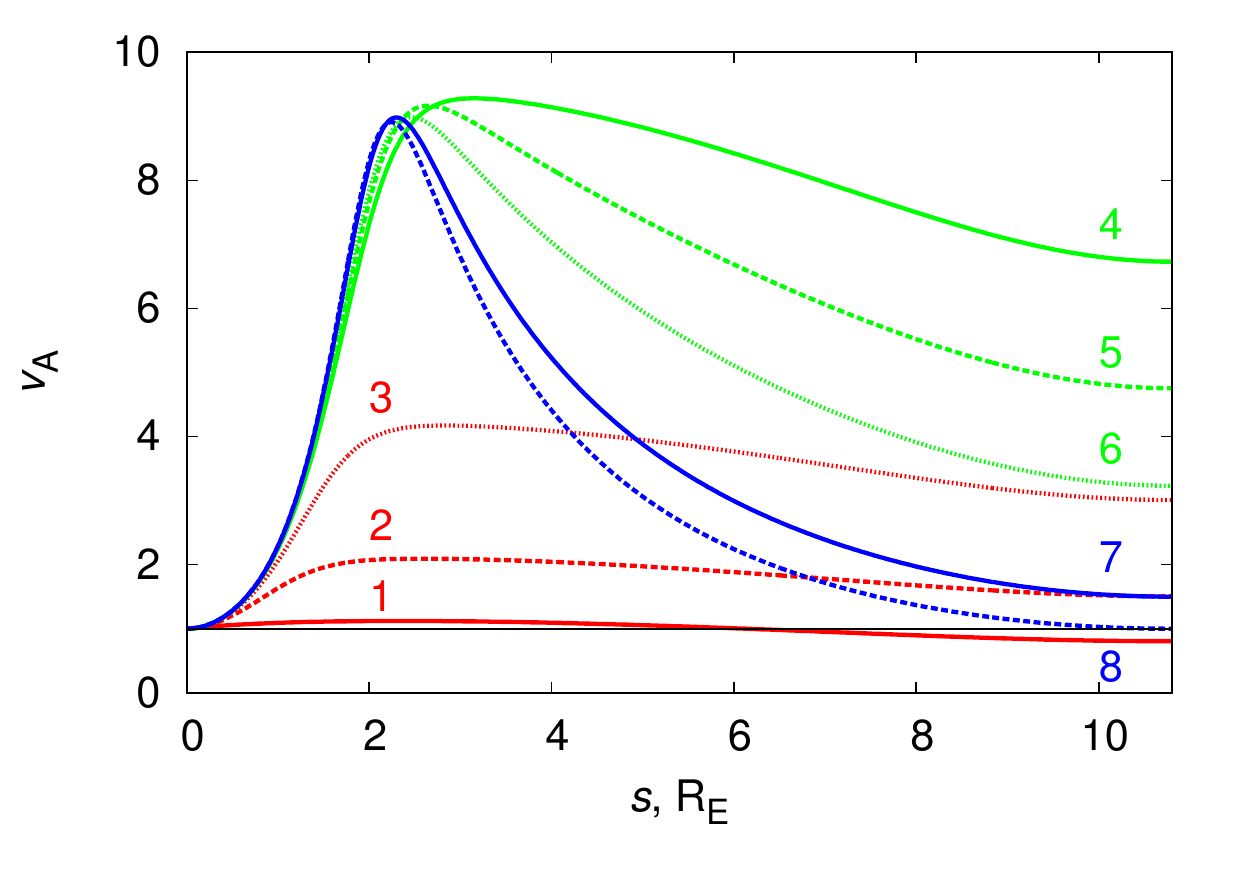}
\caption{Alfv$\acute{\rm e}$n velocity profiles 1--8 along the field line $s$, used in our simulations; the values are normalized by the average velocity $v_{\rm A}'$ (see text).}
\end{figure}

\begin{figure}[t!]
\includegraphics[width=1.0\columnwidth, bb=0 0 360 252, clip]{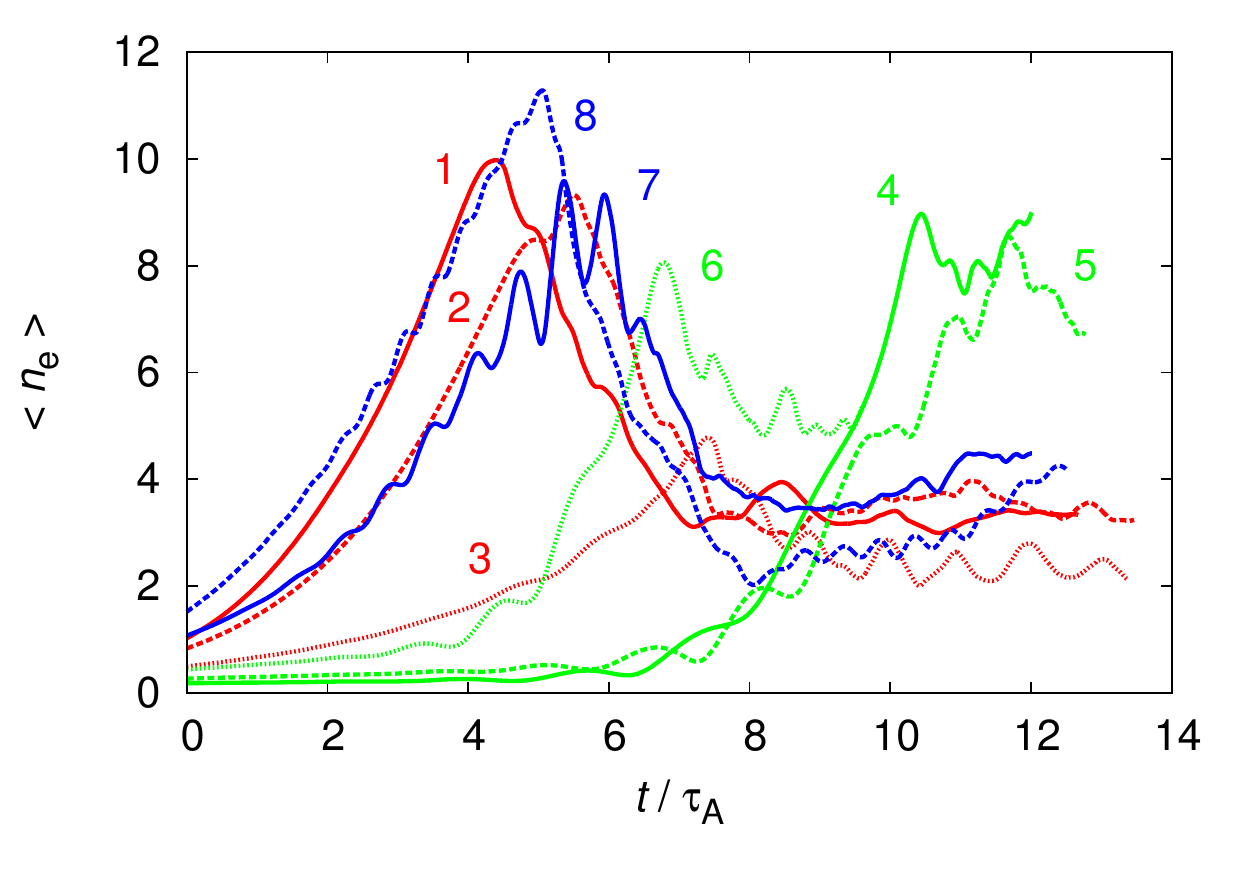}
\caption{Average electron density at the ionosphere as function of time; $\tau_{\rm A}$ is the Alfv$\acute{\rm e}$n transit time. Shown are for the eight cases of $v_{\rm A}$ profiles in Fig.\ 1.}
\end{figure}

\begin{figure*}[t]
\includegraphics[scale=0.7, bb=220 80 1000 620, clip]{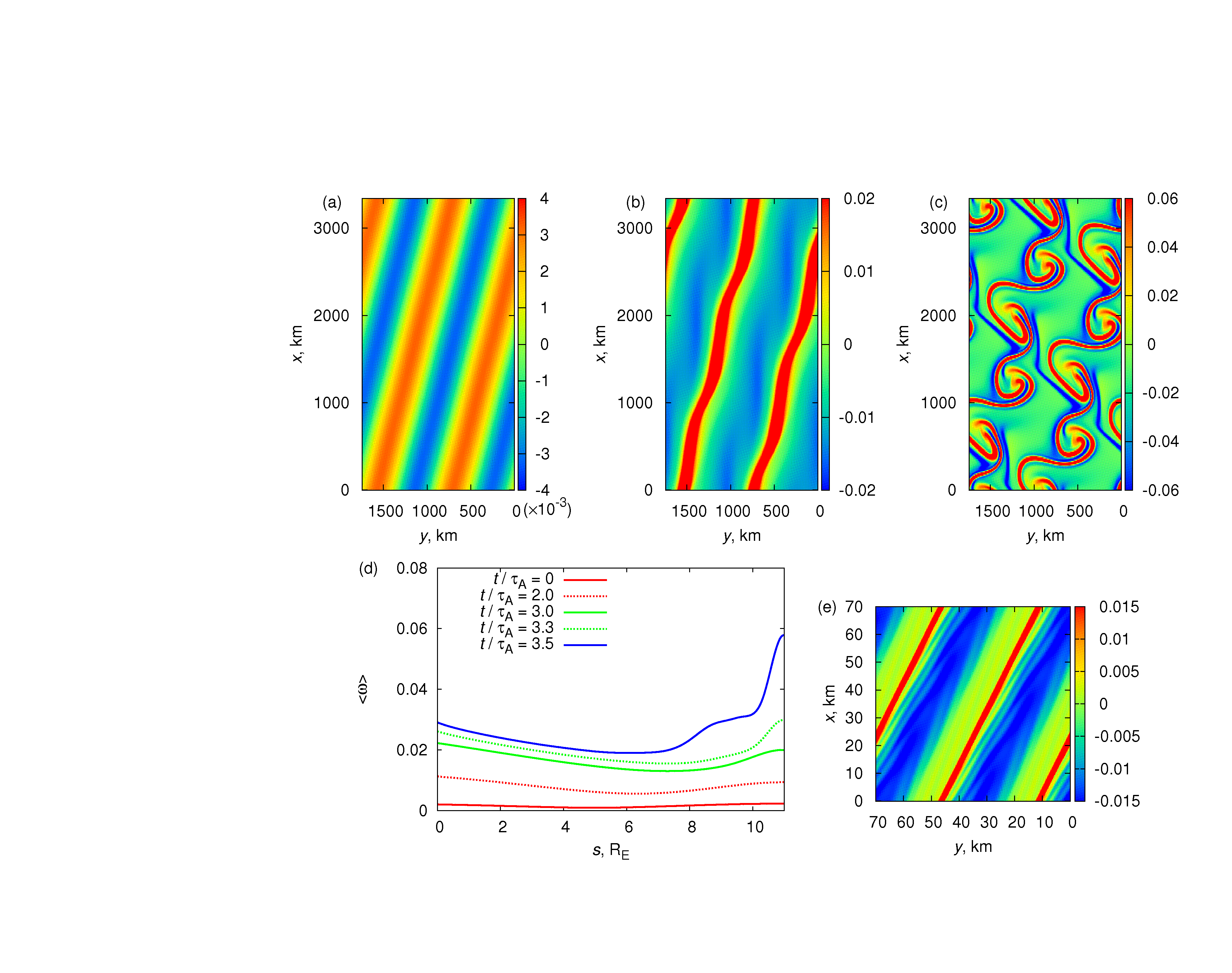}
\caption{Vorticity $\omega(x,y)$ at the magnetic equator $s=l$ at $t/\tau_{\rm A} = 0$ (a), 3 (b), and 3.5 (c), respectively, in the case 1 of $v_{\rm A}$; the values are normalized. (d) Temporal variation in the average vorticity along the field line. (e) Vorticity at the ionosphere $s=0$ at $t/\tau_{\rm A} = 3.5$.}
\end{figure*}

We used the 4th-order finite difference methods in space and time to solve Eqs.\ (1)--(4), including sharp gradients of $v_{\rm A}$, with grid numbers of (256, 256, 512) for the $\nu$, $\varphi$, and $s$ directions, respectively. The time resolution was changed in accord with the Courant condition: $\max(\mbox{\boldmath $v$}_1 / \Delta x(s)) \Delta t < 0.25$. A higher-order low-pass filter [Lele, 1992] was used along with the numerical viscosity and resistivity $\nu_{\rm v} = \eta = 1\times 10^{-7} /B_0(s)$. Regarding the calculation domain $\mbox{\boldmath $x$}_\perp(s=0) \equiv [x,y]$, $x$ and $y$ pointed southward and eastward, respectively, in the southern hemisphere. We set a periodic boundary in the $\mbox{\boldmath $x$}_\perp$ direction, e.g., at $x$, $y = 0$ and $l_\perp = 70$ km (thus $\Delta x \approx 0.27$ km) at the ionosphere $s=0$. An asymmetric boundary for the magnetic field $\psi = 0$ (or $j_\parallel = 0$) was set at the magnetic equator $s=l$. At the ionospheric boundary of $\phi$, Eq.\ (4) was solved using the multigrid-BiCGStab method. Here, values of the control parameters were referred to Hiraki [2014]; $B_0=5.7\times10^{-5}$ T, $\mu_{\rm P} / \mu_{\rm H}=0.5$, $D=4\times 10^5$ m$^2$/s, $R=2\times10^{-3}$ /s, and the ambient density $n_{\rm e0} = 3.8 \times10^4$ cm$^{-3}$ at $s=0$. In this paper, the convection electric field was set to point southward with an amplitude of $E_0 = 60$ mV/m at $s=0$, i.e., an westward drift of $v_0 \approx 1.1$ km/s. 

Different from $v_{\rm A}={\rm const}$ in our previous study [Hiraki, 2014], a field line variation in the Alfv$\acute{\rm e}$n velocity $v_{\rm A}(s) \equiv B_0(s) / \sqrt{\mu_0 m_{\rm i} n_{\rm i}(s)}$ was taken into account. Here, $\mu_0$ is the permeability, $m_{\rm i}$ the proton mass, and $n_{\rm i}$ the ion density, respectively. The modeled ion density [HW, 2011]
\begin{equation}
 \displaystyle n_{\rm i}(s) = n_{\rm a} {\rm e}^{-(\hat{r}(s) - 1)/h_{\rm a}} + n_{\rm b} \hat{r}(s)^{-q} + n_{\rm c}
\end{equation}
was used. Here, $\hat{r}$ is the normalized distance from the Earth center to the position $s$; $\hat{r}(0)=1$ and $\hat{r}(l)\approx 8.5$. We fixed $n_{\rm i}(0) = 7\times 10^5$ cm$^{-3}$, $h_{\rm a} = 1/6$, and $n_{\rm c} = 0$ cm$^{-3}$. By changing $n_{\rm b}=8.5\times10^1$--$7\times10^5$ cm$^{-3}$ and $q=2.4$--6.4, we obtained eight $v_{\rm A}$ profiles characterized by ionospheric and magnetospheric cavities shown in Fig.\ 1. These profiles were renormalized so that the Alfv$\acute{\rm e}$n transit time is fixed constant as $\tau_{\rm A} \equiv \int_0^l {\rm d}s/v_{\rm A}(s) = l/v_{\rm A}'\approx 47$ s with the average velocity $v_{\rm A}' = 1.5 \times 10^3$ km/s. 

We solved a linearized set of Eqs.\ (1)--(4) to determine the eigenfunctions $(\tilde{\phi}(s), \tilde{\psi}(s), \tilde{n}_{\rm e}(0))$ and frequency $\Omega$ of Alfv$\acute{\rm e}$n waves as functions of the perpendicular wavenumber $\mbox{\boldmath $k$}_\perp$ and the field-line harmonic number. Although field variables of an initially placed auroral arc were treated in our previous study [Hiraki, 2014], we provide only the perturbed fields $(\tilde{\phi}, \tilde{\psi}, \tilde{n}_{\rm e})$ to shed light on the pure nonlinear coupling of Alfv$\acute{\rm e}$n eigenmodes in this study. The fundamental mode with ${\rm Re}(\Omega) \approx 16$ mHz and $\mbox{\boldmath $k$}_\perp / 2 \pi = (k_x, k_y) = (1, 2)$ was found to have the maximum growth rate $\gamma \equiv {\rm Im}(\Omega)$ for the case 1 of $v_{\rm A}$ in Fig.\ 1. We performed 8 runs using each $v_{\rm A}$ in Fig.\ 1 and yielding the perturbed fields of the (1, 2) mode along with a stable $(2, -1)$ mode with $\gamma<0$; i.e., $\phi = \epsilon \tilde{\phi}_{1,2} (1 + \delta {\rm e}^{{\rm i}k_{(2,-1)} x_\perp})$ with $\epsilon = 10^{-5}$ and $\delta = 10^{-2}$ at $t = 0$, $\psi$ and $n_{\rm e}$ as well but $\delta=0$.

\section{Results}\label{sec: 3}
\subsection{Overall trend}\label{sec: 3.1}
We performed nonlinear simulations of Alfv$\acute{\rm e}$n waves using Eqs.\ (1)--(4) for 8 cases of the velocity profiles as shown in Fig.\ 1. Figure 2 shows the temporal evolution of the root mean square density $\langle n_{\rm e} \rangle$ at the ionosphere; we illustrated the density that has a slower change than the other variables, vorticity $\omega$ and current $j_\parallel$. Note that slight differences in the initial values are originated from a dependence of eigenfunction $\tilde{n}_{\rm e1,2}$ on $v_{\rm A}(s)$. As expected from the $v_{\rm A}$ profile, we can divide the results into three groups, that is, cases 1--3, cases 4--6, and cases 7, 8. 

In the cases 1 and 2 including a weak gradient of $v_{\rm A}$, the density grows with an extremely high rate ($t/\tau_{\rm A} \approx 1$--2) to a high value of $\langle n_{\rm e} \rangle = 8$--10; note that the ambient density is set as $n_{\rm e0}=10$ in this unit. It causes the linearly stable mode $(2, -1)$ to gain energies, and $\langle n_{\rm e} \rangle$ shows a saturation through a nonlinear coupling to the (1, 2) mode. The density decreases into a low level as $\langle n_{\rm e} \rangle = 2$--3 after the waves are stabilized along the field line. The cases 4--6 are characterized by a developed ionospheric cavity. Since the fundamental eigenmode has a small growth rate in these cases [HW, 2012], the density slowly increases until $t/\tau_{\rm A} \approx 8$. Though we mention later in Sec.\ \ref{sec: 3.3}, a short-wavelength wave is trapped in the cavity region, a gradual oscillation with a period of $t/\tau_{\rm A}\approx 1.6$ appears, and after that the density reaches a high level ($\langle n_{\rm e} \rangle \approx 8$) during a rapid growth of IAR modes. The cases 7 and 8, respectively, have high growth rates compared to those in the cases 2 and 1. This arises from the situation that the fundamental eigenmode, escaping from the magnetospheric cavity, can penetrate into the ionosphere without any loss of amplitude; see HW [2011] for details. The remarkable feature is a fast oscillation with a period of $t/\tau_{\rm A} \approx 0.6$ during the growth phase; waves are trapped in the ionospheric cavity in this period. After $t/\tau_{\rm A} \approx 8$, $\langle n_{\rm e} \rangle$ reaches a steady state but is a higher level ($\langle n_{\rm e} \rangle \approx 4.5$) than the cases 1--3. 

\begin{figure}[b]
\includegraphics[width=1.0\columnwidth, bb=0 0 360 252, clip]{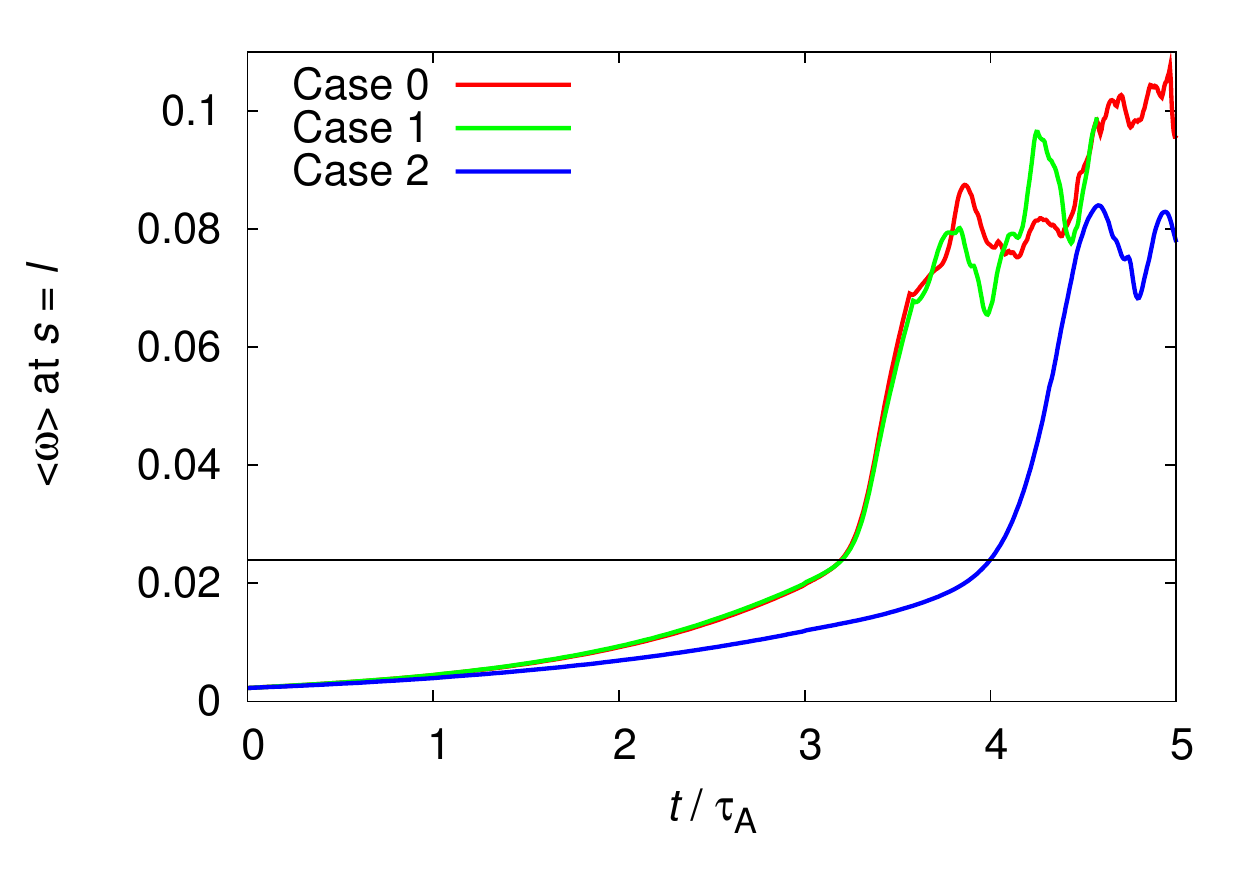}
\caption{Temporal variation in the average vorticity at the magnetic equator $s=l$, in the cases 0--2 of $v_{\rm A}$ (see text).}
\end{figure}

\subsection{Cases 1--3 of $v_{\rm A}$}\label{sec: 3.2}
Figure 3 (a)--(c) shows the temporal variation in vorticity $\omega(x,y)$ at the magnetic equator $s=l$ in the case 1; panels are for $t/\tau_{\rm A} = 0$, 3, and 3.5, respectively. Also, the average vorticity $\langle \omega \rangle(s)$ at each field line position $s$ is shown in panel (d), while $\omega(x,y)$ at the ionosphere at $t/\tau_{\rm A}=3.5$ is in panel (e). We found from Fig.\ 3 (b) and (c) that the initial eigenmode at $s=l$ strongly deforms into a thinner structure of $\omega>0$, and the vortex street forms just after that. We also found from Fig.\ 3 (d) that this secondary instability at the nonlinear stage starts around a localized area of $s=10$--11 R$_{\rm E}$; erosion, or pileup of vortices, into the lower altitude occurs shortly. On the other hand, the initial eigenmode structure keeps until $t/\tau_{\rm A}=3.5$ in the ionospheric side (e), though areas of $\omega>0$ becomes thinner. We see formation of thin hair-like structures on the both ends. 

In order to elucidate the cause of the secondary instability depicted above, we have performed one more run with $v_{\rm A}={\rm const}$ (hereafter, case 0) to be compared with the cases 1 and 2. Figure 4 shows the results of $\langle \omega \rangle(s=l)$. It is clear that there is a dramatic change of growth of the vorticity (flow) perturbations at $t/\tau_{\rm A}\approx 3.3$ for the cases 0 and 1, and at $t/\tau_{\rm A}\approx 4.1$ for the case 2. The values of $\langle \omega \rangle(s=l)\approx0.024$ and $\langle v_{1y} \rangle \approx v_0 \approx 26$ km/s at these times were estimated. It means that, when the wave perturbation offsets the background convection field, a vortex street forms by the velocity shear on both sides of the maxima of $|\omega|$. In addition, we note that there is no feature of the above-like secondary instability in the case 3, which is situated between these cases 0--2 and case 4. 

\begin{figure}[h]
\includegraphics[scale=0.7, bb=80 80 700 580, clip]{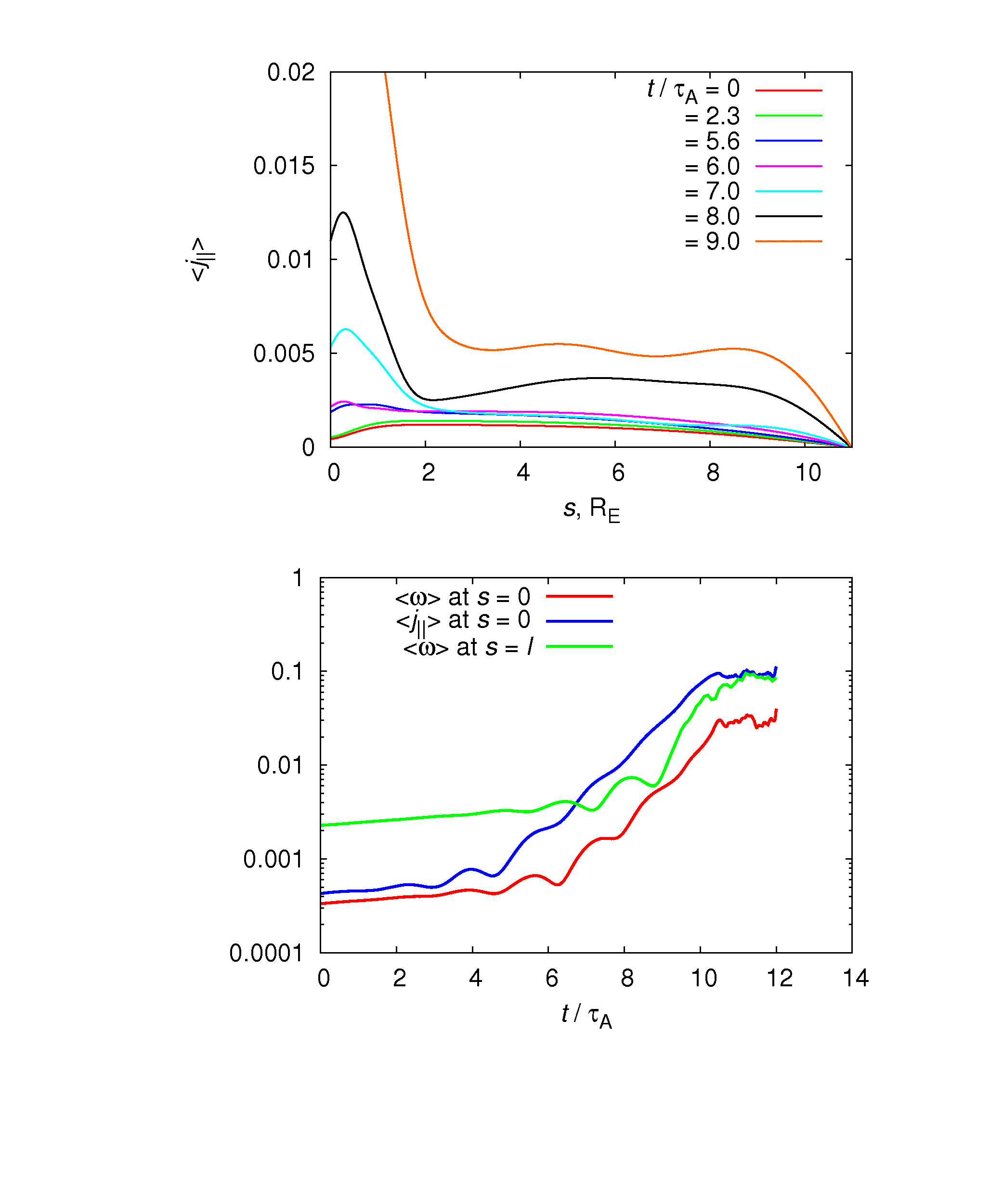}
\caption{Temporal variation in the average field-aligned current along the field line (upper panel), in the case 4 of $v_{\rm A}$. Average vorticity and current at $s=0$ and vorticity at $s=l$ as function of time (bottom panel).}
\end{figure}

\begin{figure}[b]
\includegraphics[scale=0.75, bb=100 80 700 730, clip]{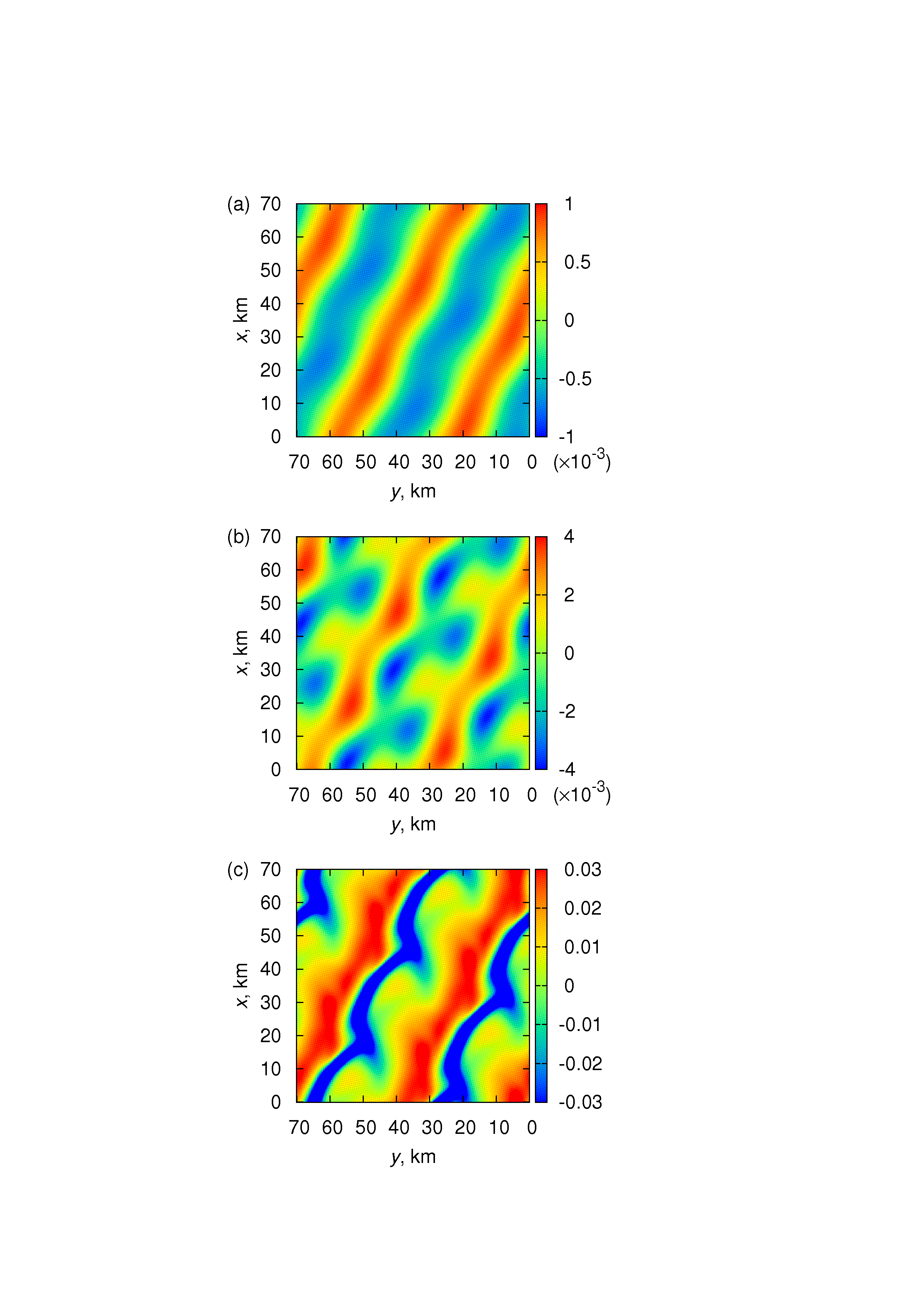}
\caption{Field-aligned current $j_\parallel(x,y)$ at the ionosphere $s=0$ at $t/\tau_{\rm A}=2.3$ (a), 5.6 (b), and 9 (c), respectively, in the case 4 of $v_{\rm A}$; the values are normalized by $j' = 660$ $\mu$A/m$^2$.}
\end{figure}

\subsection{Cases 4--6 of $v_{\rm A}$}\label{sec: 3.3}
Next, let us see the behaviors of $\omega$ and $j_\parallel$ in the case 4 where only the ionospheric cavity is deepened. Figure 5 shows the temporal variation in the average field-aligned current $\langle j_\parallel \rangle(s)$ (upper panel), besides, in $\langle \omega \rangle$ and $\langle j_\parallel \rangle$ at $s=0$, and $\langle \omega \rangle$ at $s=l$ (lower panel). The upper panel reveals that the wave magnetic perturbation, or $\langle j_\parallel \rangle$, begins to be trapped in the ionospheric cavity region $s=0$--2 R$_{\rm E}$ at $t/\tau_{\rm A}=2.3$. We also find from the lower panel that $\langle j_\parallel \rangle(s=0)$ increases and has a weak oscillation that means incidence and reflection of waves at the ionosphere. An oscillatory behavior of $\langle \omega \rangle(s=0)$ in phase with $\langle j_\parallel \rangle(s=0)$ occurs at $t/\tau_{\rm A}=4$, and these show a faster growth than in the previous period. The vorticity $\langle \omega \rangle(s=l)$ also oscillates from $t/\tau_{\rm A}=6$ and grows to be a high level until $t/\tau_{\rm A}\approx10$ since a moderate-amplitude wave propagates from the ionosphere. 

Figure 6 shows the temporal variation in field-aligned current $j_\parallel(x,y)$ at $s=0$ at $t/\tau_{\rm A}=2.3$, 5.6, and 9. We find that the initially given mode becomes wavy at $t/\tau_{\rm A}=2.3$, which indicates wave trapping in the ionospheric cavity as mentioned above. High wavenumber modes in $j_\parallel$ appear, involving a growth of $\langle \omega \rangle(s=0)$ (see Fig.\ 5) at $t/\tau_{\rm A}=5.6$. It proves that the first harmonic eigenmode is generated from the fundamental perturbation due to a slight phase difference in $\omega$, $j_\parallel$, and $n_{\rm e}$. The initial structure of $j_\parallel>0$ strongly deforms, and small patches are newly produced at each left-upper side at $t/\tau_{\rm A}=9$. The field-aligned current reaches a high value of $\langle j_\parallel \rangle \approx 20$ $\mu$A/m$^2$ at this time. 

\begin{figure}[t]
\includegraphics[width=1.0\columnwidth, bb=0 0 360 252, clip]{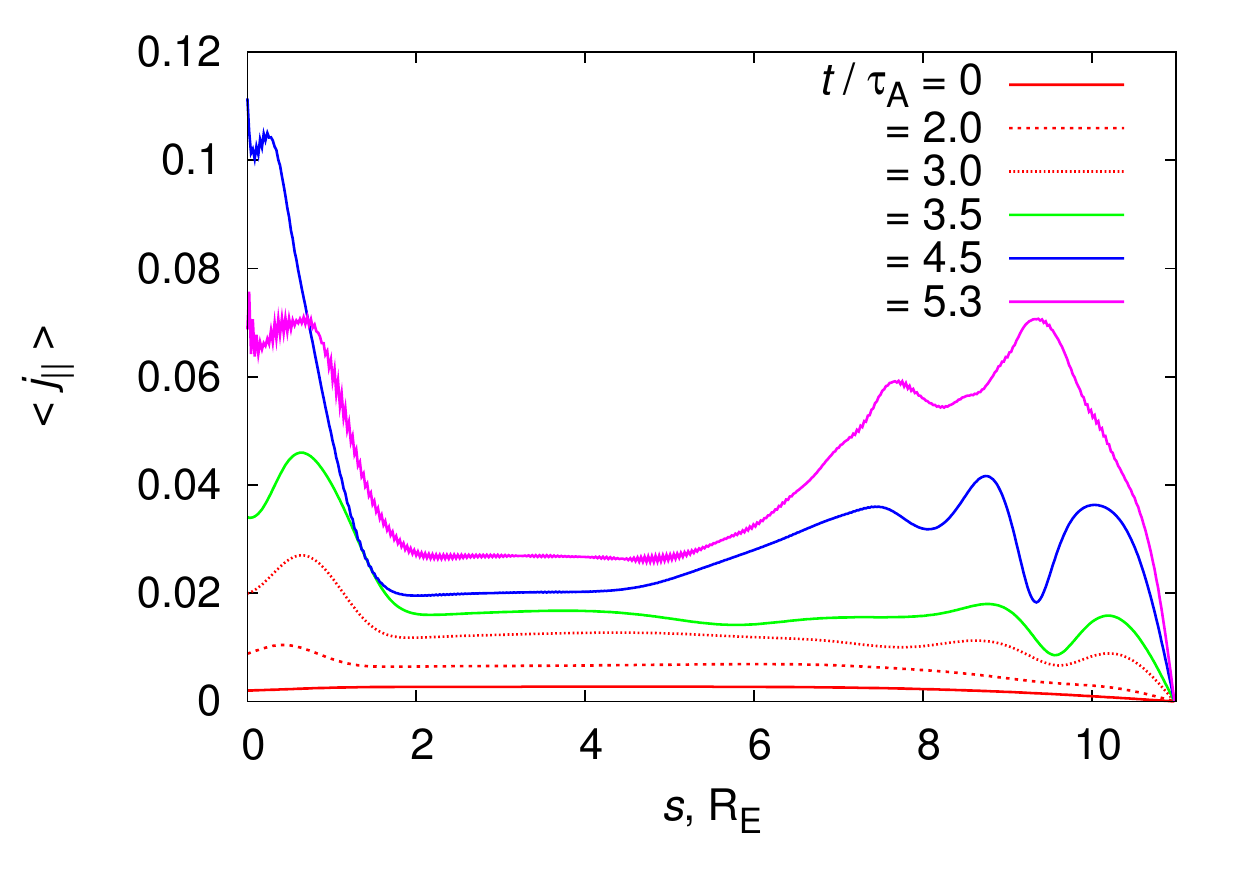}
\caption{Temporal variation in the average field-aligned current along the field line, in the case 7 of $v_{\rm A}$.}
\end{figure}

\begin{figure*}[t]
\includegraphics[scale=0.67, bb=200 80 1000 600, clip]{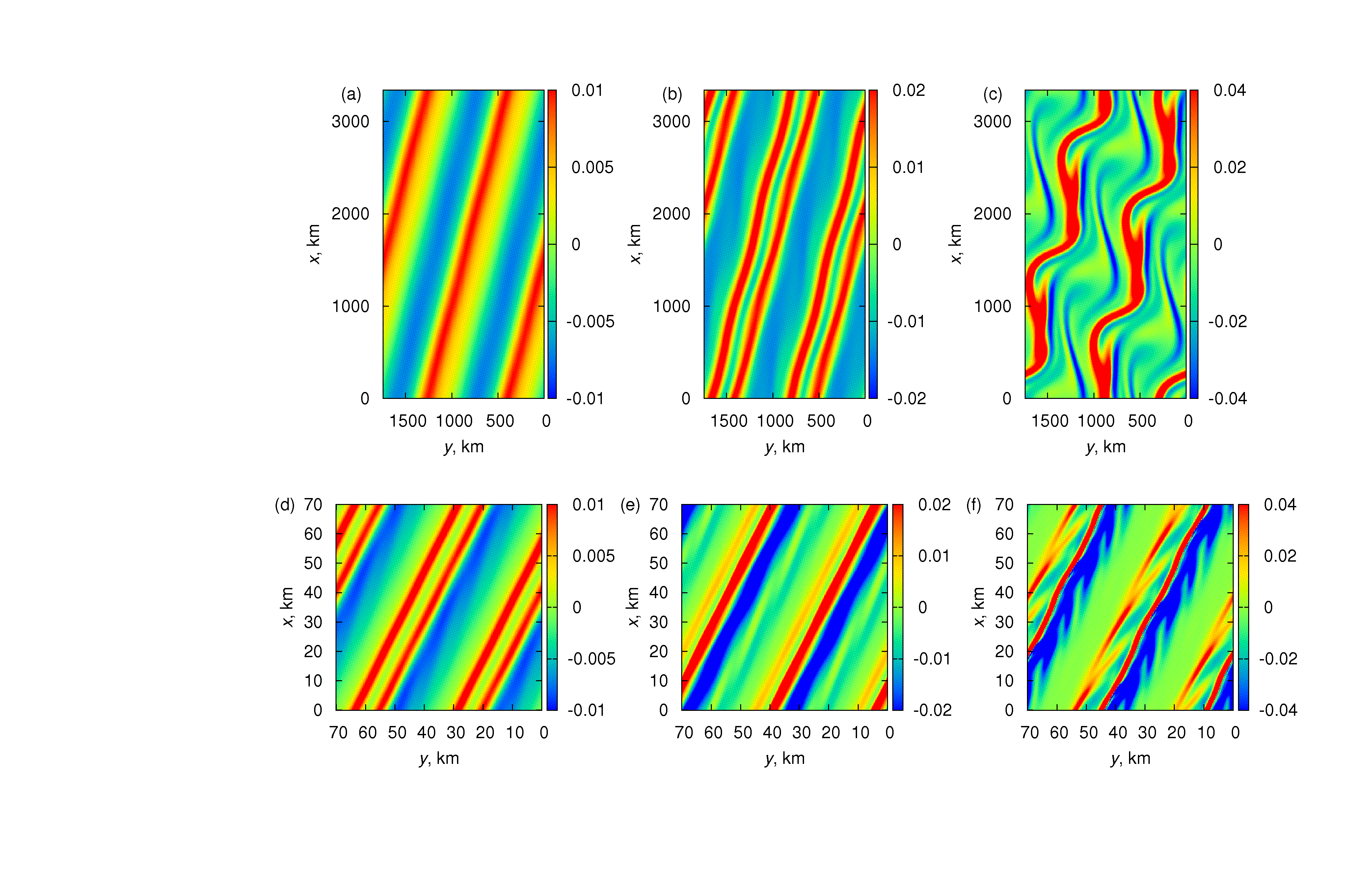}
\caption{Vorticity $\omega(x,y)$ at the magnetic equator $s=l$ (a--c) and the ionosphere $s=0$ (d--f) at $t/\tau_{\rm A} = 2$, 3.5, and 4.5, respectively, in the case 7 of $v_{\rm A}$.}
\end{figure*}

\subsection{Cases 7 and 8 of $v_{\rm A}$}\label{sec: 3.4}
Let us finally see the behaviors of variables in the case 7 where there is a unique oscillation in the linear stage (see Fig.\ 2). Figure 7 shows the temporal variation in $\langle j_\parallel \rangle(s)$. We find that an initially placed eigenfunction quickly deforms at $t/\tau_{\rm A} = 2$, and the ionospheric Alfv$\acute{\rm e}$n resonance (IAR) occurs. The half-wavelength IAR wave oscillates, while a wavy structure forms at $s=7$--11 R$_{\rm E}$, at $t/\tau_{\rm A} = 3$--3.5. The latter means that part of IAR waves, through their ionospheric feedback coupling, is shot to the magnetic equator. The one-wavelength IAR wave is produced at $s=0$ at $t/\tau_{\rm A} = 4.5$. A rapid growth of $\langle j_\parallel \rangle(s)$ up to 70 $\mu$A/m$^2$ in the cavity region is found. The current amplifies until $t/\tau_{\rm A} = 5.3$ in the magnetospheric side, extending down to $s \approx 5$ R$_{\rm E}$; pileup of vortices occurs in the background. It is clear that the hybrid Alfv$\acute{\rm e}$n resonant (HAR) waves [HW, 2012] grow. 

Figure 8 shows the temporal variation in vorticity $\omega(x,y)$ at the magnetic equator $s=l$ (a--c) and the ionosphere $s=0$ (d--f) at $t/\tau_{\rm A} = 2$, 3.5, and 4.5. The first striking change appears in the $s=0$ side, that is, splitting of initial arcs along with thin hair-like structures on the poleward side. This feature is related to the trapping of IAR waves depicted in Fig.\ 7. The effect of this splitting is transmitted (as part of IAR waves) to the $s=l$ side until $t/\tau_{\rm A} = 3.5$, and a vortex structure forms at $t/\tau_{\rm A} = 4.5$. At the same time, a distinct structure, or an oblique slice forms in the initial arcs at $s=0$. We clarified the timing that vortices appear at $s=l$. The same as the cases 0--2, vorticity and the $y$-component of perturbed velocity at $s=l$ reach $\langle \omega \rangle(s=l)\approx0.024$ and $\langle v_{1y} \rangle \approx v_0 \approx 26$ km/s, respectively, at $t / \tau_{\rm A} \approx 4.3$. Vortices in this case are slightly distorted compared to the case 1 shown in Fig.\ 3, because their formation is strongly affected by IAR waves propagating from the ionosphere. However, we emphasize that the secondary nonlinear instability at $s=l$ is triggered by a flow shear formed after the wave perturbation offsets the background convection field. It is also marvelous that the total perturbed velocity peaks at $\langle v_{1} \rangle_{\max} \approx 65$ km/s and $|v_{1}|_{\max} \approx 220$ km/s at the latter period $t/\tau_{\rm A} = 6.2$.

\section{Discussion}\label{sec: 4}
Let us summarize the effects of the Alfv$\acute{\rm e}$n velocity cavities on the wave growth obtained above. For the case 0 with $v_{\rm A}(s)={\rm const}$ and a dipole $B_0(s)$, we find that the convection drift is offset through a growth of perturbed velocity fields and a secondary instability occurs at the magnetic equator. The similar behavior was found in the simpler slab geometry of $B_0 = {\rm const}$ [Watanabe, 2010]. Much simpler system that describes shear Alfv$\acute{\rm e}$n wave dynamics is set up by removing the ionosphere; however, it provides an almost self-evident solution, or a uniform propagation of given waves. We call the case 0 the minimal model of shear Alfv$\acute{\rm e}$n waves developed in the MI coupling system. Only the growth of internal waves on the flux tube of our interest, without any external force, produces a strong flow perturbation at the magnetic equator. This is originated from the fact that the fundamental eigenmode has a wave form of the same order of amplitude throughout the field line. The maximum of the flow perturbations at $s=l$ in this case is estimated to be $|v_{1}|_{\max} \approx 230$ km/s at $t/\tau_{\rm A} = 5$ and the $x$-component is one third of that, which is similar to the case 7 shown in Sec.\ \ref{sec: 3.4}. So-called "bursty flow" that occurs in substorm onsets has been known to have a magnitude of $v_x = 100$--300 km/s [Lee et al., 2012; Sergeev et al., 2012]. We suppose that the flow perturbation obtained in this study could not be the bursty flow itself but be the first indication of it. 

Next, we consider what knowledge is given by the expansion from cases 0--2 to cases 4--5. That is that trapping of IAR waves occurs after the fundamental mode structure deforms. This is originated from the fact that the first harmonic eigenmode has a wave form of which the half wavelength is trapped in the cavity region (see Fig.\ 5). It is marvelous that, when the IAR modes grow, the secondary instability (vortex formation) at the magnetic equator side is suppressed. When $\omega(s=l)$ becomes large after $t/\tau_{\rm A}\approx 10$ as shown in Fig.\ 5, structures are surely changed to be turbulent. By the next expansion to cases 7 and 8, we clarified that a high-amplitude wave is trapped in the cavity side, grows, and the field energy is transported to the magnetic equator side. It causes the flow shear instability. It means that these cases combine the essences of the above two groups (0--2 and 4--5). We emphasize that the first harmonic wave is excited in the cavity region, and then the second- and the third harmonic waves are excited to form the hybrid modes, resulting in the strong flow perturbation at the magnetic equator. It brings a directivity "from one side (cavity) to the other side (magnetic equator)" into our system. The existence of the directivity is supported by observations of AKR [Morioka et al., 2010]. They claimed that an enhancement of the field-aligned current occurs at the acceleration region (nearly equals to the height of $v_{\rm A}$ peaks) at substorm onsets, and then it causes the enhancement of flow fields at the magnetic equator. 

When the background cavity becomes deep, a nonlinear coupling between the stable and unstable fundamental modes is promoted due to a perpendicular phase shift of $n_{\rm e}$ and $\phi$. The conversion from the initially placed fundamental mode (cases 1--3) to the IAR and HAR waves ($\parallel \mbox{\boldmath $B$}_0$) occurs simultaneously. This property is seen in the values of $\langle n_{\rm e} \rangle$, especially those in the quasi-steady states after saturation. The values in the cases 4--6 (IAR) are largest and those in the cases 7 and 8 (HAR) are the next one at $t/\tau_{\rm A} \ge 8$. On the other hand, the fundamental mode in the cases 1--3 is considered not to bring any strong energy in the ionosphere once it is stabilized. If we relate our result "the vortices at the magnetic equator associated with IAR wave propagation" to the evolution of auroral structures, we need to investigate the relationship between the vortices and the field-aligned electric field $E_\parallel$. By doing a test particle simulation under the parallel field as done by Chaston et al.\ [2006], our next problem is to clarify which physical quantity directly concerns the auroral luminosity. Although there has been many theoretical studies [Chaston et al., 2008; Watt et al., 2009; Lysak and Song, 2008] on the electron acceleration by Alfv$\acute{\rm e}$n waves, a 3D nonlinear calculation that treats the interaction of shear Alfv$\acute{\rm e}$n and inertial Alfv$\acute{\rm e}$n waves along a full field line has not been implemented. For the realistic setup, we should adopt a deeper cavity by one order than that in the case 8 of this study: $v_{\rm A, \min}\approx 10^6$ m/s to $v_{\rm A,\max} \approx 10^8$ m/s. The IAR and HAR modes could be generated to play a major role in auroral structuring and energy transport in the MI coupling system.

\section{Conclusion}
We performed MHD simulations of shear Alfv$\acute{\rm e}$n waves in a full field line system with MI coupling and Alfv$\acute{\rm e}$n velocity cavities. 
Feedback instability of the Alfv$\acute{\rm e}$n resonant modes showed various nonlinear features: i) a secondary flow shear instability occurs at the magnetic equator, ii) trapping of the ionospheric Alfv$\acute{\rm e}$n resonant modes facilitates deformation of auroral fine structures, and iii) waves emitted from the ionospheric cavity cause vortices and magnetic oscillations around the magnetic equator side. 
Essential features in the initial brightening of auroral arc at substorm onsets could be explained by growth of the Alfv$\acute{\rm e}$n resonant modes, which are the nature of the field line system responding to a rapid change in the background conditions. 
The IAR and HAR modes could play a major role in auroral structuring and energy transport in the MI coupling system.


\newpage 

\begin{thebibliography}{}
\bibitem{}
J.\ C.\ Samson, D.\ D.\ Wallis, T.\ J.\ Hughes, F.\ Creutzberg, J.\ M.\ Ruohoniemi, and R.\ A.\ Greenwald (1992) {\it J.\ Geophys.\ Res., 97,} 8495. 
\bibitem{}
R.\ L.\ Lysak, and Y.\ Song (2008) {\it Geophys.\ Res.\ Lett., 35,} L20101, doi:10.1029/2008GL035728.
\bibitem{}
I.\ J.\ Rae, I.\ R.\ Mann, K.\ R.\ Murphy, D.\ K.\ Milling, A.\ Parent, V.\ Angelopoulos, H.\ U.\ Frey, A.\ Kale, C.\ E.\ J.\ Watt, S.\ B.\ Mende et al.\ (2009) {\it J.\ Geophys.\ Res., 114,} A00C09, doi:10.1029/2008JA013559. 
\bibitem{}
R.\ Rankin, J.\ Y.\ Lu, R.\ Marchand, and E.\ F.\ Donovan (2004) {\it Phys.\ Plasmas, 11,} 1268--1276. 
\bibitem{}
V.\ G$\acute{\rm e}$not, F.\ Mottez, and P.\ Louarn (2000) {\it J.\ Geophys.\ Res., 105,} 27,611. 
\bibitem{}
C.\ C.\ Chaston, J.\ W.\ Bonnell, C.\ W.\ Carlson, J.\ P.\ McFadden, R.\ E.\ Ergun, and R.\ J.\ Strangeway (2003) {\it J.\ Geophys.\ Res., 108(A4),} 8003. 
\bibitem{}
G.\ M.\ Erickson, N.\ C.\ Maynard, W.\ J.\ Burke, G.\ R.\ Wilson, and M.\ A.\ Heinemann (2000) {\it J.\ Geophys.\ Res., 105,} 25,265. 
\bibitem{}
C.\ W.\ Carlson, R.\ F.\ Pfaff, and J.\ G.\ Watzin (1998) {\it Geophys.\ Res.\ Lett., 25(12),} 2013--2016. 
\bibitem{}
S.\ R.\ Hebden, T.\ R.\ Robinson, D.\ M.\ Wright, T.\ Yeoman, T.\ Raita, and T.\ B$\ddot{\rm o}$singer (2005) {\it Ann.\ Geophys., 23,} 1711--1721. 
\bibitem{}
V.\ G$\acute{\rm e}$not, P.\ Louarn, and F.\ Mottez (2004) {\it Ann.\ Geophys., 22,} 2081--2096. 
\bibitem{}
C.\ C.\ Chaston, V.\ G$\acute{\rm e}$not, J.\ W.\ Bonnell, C.\ W.\ Carlson, J.\ P.\ McFadden, R.\ E.\ Ergun, R.\ J.\ Strangeway, E.\ J.\ Lund, and K.\ J.\ Hwang (2006) {\it J.\ Geophys.\ Res., 111(A3),} A03206. 
\bibitem{}
A.\ Morioka, Y.\ Miyoshi, Y.\ Miyashita, Y.\ Kasaba, H.\ Misawa, F.\ Tsuchiya, R.\ Kataoka, A.\ Kadokura, T.\ Mukai, K.\ Yumoto et al.\ (2010) {\it J.\ Geophys.\ Res., 115,} A11213. 
\bibitem{}
$\O$.\ Holter, C.\ Altman, A.\ Roux, S.\ Perraut, A.\ Pedersen, H.\ P$\acute{\rm e}$cseli, B.\ Lybekk, J.\ Trulsen, A.\ Korth, and G.\ Kremser (1995) {\it J.\ Geophys.\ Res., 100(A10),} 19109--19119, doi:10.1029/95JA00990. 
\bibitem{}
M.\ Y.\ Park, D.-Y.\ Lee, S.\ Ohtani, and K.\ C.\ Kim (2010) {\it J.\ Geophys.\ Res., 115,} A11203, doi:10.1029/2010JA015566. 
\bibitem{}
T.-H.\ Watanabe (2010) {\it Phys.\ Plasmas, 17,} 022904, doi:10.1063/1.3304237. 
\bibitem{}
Y.\ Hiraki, and T.-H.\ Watanabe (2011) {\it J.\ Geophys.\ Res., 116,} A11220, doi:10.1029/2011JA016721. 
\bibitem{}
Y.\ Hiraki, and T.-H.\ Watanabe (2012) {\it Phys.\ Plasmas, 19,} 102904, doi: 10.1063/1.4759016.
\bibitem{}
V.\ Chmyrev, S.\ Bilichenko, O.\ A.\ Pokhotelov, V.\ A.\ Marchenko, V.\ Lazarev, A.\ Streltsov, and L.\ Stenflo (1988) {\it Phys.\ Scr., 38,} 841, 1988. 
\bibitem{}
Y.\ Hiraki (2014) arXiv:1408.5225 [physics.space-ph]. 
\bibitem{}
S.\ A.\ Lele (1992) {\it J.\ Comput.\ Phys., 103,} 16--42. 
\bibitem{}
D.-Y.\ Lee, H.-S.\ Kim, S.\ Ohtani, and M.\ Y.\ Park (2012) {\it J.\ Geophys.\ Res., 117,} A01207, doi:10.1029/2011JA017246.
\bibitem{}
V.\ A.\ Sergeev, V.\ Angelopoulos, and R.\ Nakamura (2012) {\it Geophys.\ Res.\ Lett., 39,} L05101, doi:10.1029/ 2012GL050859. 
\bibitem{}
C.\ C.\ Chaston C.\ Salem, J.\ W.\ Bonnell, C.\ W.\ Carlson, R.\ E.\ Ergun, R.\ J.\ Strangeway, and J.\ P.\ McFadden (2008) {\it Phys.\ Rev.\ Lett., 100,} 175003. 
\bibitem{}
C.\ E.\ J.\ Watt, and R.\ Rankin (2009) {\it Phys.\ Rev.\ Lett., 102,} 045002. 

\end{thebibliography}


\end{document}